\documentclass[aps,prl,twocolumn,showpacs,superscriptaddress]{revtex4}
\usepackage{amsmath}
\usepackage{graphicx,color}% Include figure files

\newcommand{\nn}{\nonumber\\}
\newcommand{\ba}{\begin{eqnarray}}
\newcommand{\ea}{\end{eqnarray}}
\newcommand{\la}[1]{\label{#1}}

\begin{document}

%\preprint{}

\title{Effects of jet-induced medium excitation in $\gamma$-hadron correlation in A+A collisions }



%\email[]{}
%\thanks{}
\author{Wei Chen}
\affiliation{Key Laboratory of Quark and Lepton Physics (MOE) and Institute of Particle Physics, Central China Normal University, Wuhan 430079, China}
\author{Shanshan Cao}
\affiliation{Nuclear Science Division MS 70R0319, Lawrence Berkeley National Laboratory, Berkeley, California 94720}
\affiliation{Department of Physics and Astronomy, Wayne State University, Detroit, Michigan 48201}
\author{Tan Luo}
\affiliation{Key Laboratory of Quark and Lepton Physics (MOE) and Institute of Particle Physics, Central China Normal University, Wuhan 430079, China}
\author{Long-Gang Pang}
\affiliation{Physics Department, University of California, Berkeley, California 94720}
\affiliation{Nuclear Science Division MS 70R0319, Lawrence Berkeley National Laboratory, Berkeley, California 94720}
%\author{Horst Stoecker}
%\affiliation{Frankfurt Institute for Advanced Studies, Ruth-Moufang-Strasse 1, 60438 Frankfurt am Main, Germany}
%\affiliation{Gesellschaft f{\" u}r Schwehrionenforschung, Planckstr. 1, Darmstadt, Germany}
\author{Xin-Nian Wang}
\affiliation{Key Laboratory of Quark and Lepton Physics (MOE) and Institute of Particle Physics, Central China Normal University, Wuhan 430079, China}
\affiliation{Nuclear Science Division MS 70R0319,  Lawrence Berkeley National Laboratory, Berkeley, California 94720}
\affiliation{Physics Department, University of California, Berkeley, California 94720}

%\affiliation{Key Laboratory of Quark and Lepton Physics (MOE) and Institute of Particle Physics, Central China Normal University, Wuhan 430079, China}


\date{\today}

\begin{abstract}
Coupled Linear Boltzmann Transport and hydrodynamics (CoLBT-hydro) is developed for co-current and event-by-event simulations of jet transport and jet-induced medium excitation (j.i.m.e.) in high-energy heavy-ion collisions. This is made possible by a GPU parallelized (3+1)D hydrodynamics that has a source term from the energy-momentum deposition by propagating jet shower partons and provides real time update of the bulk medium evolution for subsequent jet transport. Hadron spectra in $\gamma$-jet events of A+A collisions at RHIC and LHC are calculated for the first time that include hadrons from both the modified jet and j.i.m.e..  CoLBT-hydro describes well experimental data at RHIC on the suppression of leading hadrons due to parton energy loss. It also predicts the enhancement of soft hadrons from j.i.m.e. The onset of soft hadron enhancement occurs at a constant transverse momentum due to the thermal nature of soft hadrons from j.i.m.e. which also have a significantly broadened azimuthal distribution relative to the jet direction. Soft hadrons in the $\gamma$ direction are, on the other hand, depleted due to a diffusion wake behind the jet.

\end{abstract}

\pacs{25.75.-q, 25.75.Bh,25.75.Cj,25.75.Ld}
%\keywords{}

\maketitle

{\it Introduction.} 
 Parton energy loss in dense medium was predicted to lead to jet quenching in heavy-ion collisions \cite{Wang:1991xy}. Experimental discovery of jet quenching at Relativistic Heavy-ion Collider (RHIC) provides important evidence for the formation of strongly coupled quark-gluon plasma (QGP) \cite{Adcox:2001jp,Wang:2004dn}. Recent theoretical and experimental progress at both RHIC and the Large Hadron Collider (LHC) have advanced the jet tomography as a powerful tool for the study of QGP properties \cite{Majumder:2010qh,Qin:2015srf,Chang:2015hqa}. 

Jet quenching leads to suppression of leading hadrons, dihadron and $\gamma$-hadron correlations, due to parton 
energy loss \cite{Wang:1996yh,Vitev:2002pf,Zhang:2007ja,Zhang:2009rn}. It also modifies jet spectra, dijet and $\gamma$-jet correlations, jet profiles and jet fragmentation functions \cite{Vitev:2009rd,Aad:2010bu,Qin:2010mn,CasalderreySolana:2010eh,Young:2011qx,Dai:2012am,Chien:2015hda,Milhano:2015mng,Chang:2016gjp,Blaizot:2015lma,Kang:2017frl}. Since jets are reconstructed from collimated cluster of hadrons within a jet cone, the final jet modification will be determined not only by energy loss of the leading jet shower partons but also how the lost energy is redistributed in the medium through induced radiation, rescattering and jet-induced medium excitation (j.i.m.e.) \cite{Wang:2013cia,Blaizot:2013hx,Apolinario:2012cg,Fister:2014zxa,Casalderrey-Solana:2015vaa}. 
Similarly, dissipation of lost energy in medium can also influence soft hadron spectra associated with hard jet production \cite{Andrade:2014swa}.
Tomography of QGP with jets and jet-hadron correlations therefore requires a complete understanding of both jet transport in a fluctuating and dynamically evolving QGP medium and j.i.m.e.. 
%Studies of the hydrodynamic medium response to jet propagation can also provide information on transport properties of QGP.

Jet-induced medium excitation in heavy-ion collisions has been the subject of many recent studies \cite{CasalderreySolana:2004qm,Stoecker:2004qu,Li:2010ts}. Theoretical tools used include parton transport \cite{Xu:2004mz,Schenke:2009gb,Zapp:2008gi,Zapp:2012ak,He:2015pra,Cao:2016gvr,Cao:2017hhk}, hydrodynamics
%with a given source term for the energy-momentum deposition 
\cite{Betz:2008ka,Qin:2009uh,Tachibana:2015qxa,Tachibana:2017syd} and AdS/CFT \cite{Gubser:2007ga,Chesler:2007sv}. The Linear Boltzmann Transport (LBT) model has been developed for the study of both jet transport and j.i.m.e. in QGP \cite{Wang:2013cia,Li:2010ts,He:2015pra,Cao:2016gvr,Cao:2017hhk}. It simulates the propagation of not only jet shower partons and radiated gluons but also recoil and back reaction partons from jet-medium interaction within perturbative QCD (pQCD). 
%The LBT model neglects interactions among recoiled partons when the j.i.m.e. is only a small perturbation and applies lowest-order 
%(LO) pQCD to parton interactions at all energy scale. The hydrodynamic model, on the other hand, takes energy-momentum deposition from jet medium 
%interaction at all energy scale as a source term in the hydrodynamic equations and evaluate the hydrodynamic response of the medium to the jet-medium interaction. 

We have recently developed the first Coupled Linear Boltzmann Transport and hydrodynamics (CoLBT-hydro) in which LBT for jet propagation is coupled to (3+1)D relativistic hydrodynamics in real time. In this coupled approach, LBT provides a source term for energy-momentum deposition by propagating partons in the hydrodynamics which in turn updates the bulk medium profile for LBT in the next time step. CoLBT-hydro for event-by-event simulations is made possible only with a Graphics-Processing-Unit (GPU) parallelized (3+1)D hydrodynamics. It combines the pQCD approach for the propagation of energetic jet shower partons with the hydrodynamic evolution of the strongly coupled QGP medium, including j.i.m.e.. It therefore can describe both high and low $p_T$ phenomena in high-energy heavy-ion collisions. We report in this Letter our first study of $\gamma$-hadron correlations with CoLBT-hydro. We study in particular the effect of j.i.m.e. in soft hadrons associated with the suppression of leading hadrons due to parton energy loss in $\gamma$-jet events of heavy-ion collisions.

%\vspace{-0.1cm}
{\it LBT model.}  
In LBT model, jet transport is simulated according to a linear  Boltzmann equation \footnote{Note that in the first publication of LBT model \cite{He:2015pra}, the degeneracy factor $\gamma_b$ and an overall factor 1/2 are missing in Eq. (1) . The degeneracy factor is also missing in the formulae for scattering rate in Eqs. (4) and (9). These are all typos in the manuscript.}
%\cite{note},
\ba
p_a\cdot\partial f_a&=&\int \prod_{i=b,c,d}d[p_i] \frac{\gamma_b}{2}(f_cf_d-f_af_b)|{\cal M}_{ab\rightarrow cd}|^2
\nn && \hspace{-0.5in}\times
S_2(\hat s,\hat t,\hat u)(2\pi)^4\delta^4(p_a\!+\!p_b\!-\!p_c\!-\!p_d)+ {\rm inelastic},
\label{bteq}
\ea
where $d[p_i]=d^3p_i/[2E_i(2\pi)^3]$, $\gamma_b$ is the color-spin degeneracy for parton $b$, $f_i=1/(e^{p_i\cdot u/T}\pm1)$ $(i=b,d)$ are parton phase-space distributions in a thermal medium with local temperature $T$ and fluid velocity $u=(1, \vec{v})/\sqrt{1-\vec{v}^2}$, $f_i=(2\pi)^3\delta^3(\vec{p}-\vec{p_i})\delta^3(\vec{x}-\vec{x_i}-\vec{v_i}t)$ $(i=a,c)$ are the phase-space density for jet shower partons before and after scattering.  $S_2(\hat s, \hat t, \hat u) = \theta(\hat s\ge 2\mu_{D}^2)\theta(-\hat s+\mu_{D}^2\le \hat t\le -\mu_{D}^2$) is introduced \cite{Auvinen:2009qm} to regulate the collinear divergency  in the leading-order (LO) elastic scattering amplitude $|{\cal M}_{ab\rightarrow cd}|^2$ \cite{Eichten:1984eu}, where $\hat s$, $\hat t$, and $\hat u$ are Mandelstam variables, and $\mu_{D}^2 = 3g^2 T^2/2$ is the Debye screen mass with 3 quark flavors. The cross section of corresponding elastic collision is $d\sigma_{ab\rightarrow cd}/d\hat t=|{\cal M}_{ab\rightarrow cd}|^2/16\pi \hat s^2$. The strong coupling constant $\alpha_s=g^{2}/4\pi$ is fixed and will be fitted to experimental data.

The inelastic process of induced gluon radiation accompanying each elastic scattering is also included in LBT. The radiated gluon spectrum is simulated according to the high-twist approach \cite{Guo:2000nz,Wang:2001ifa},
\ba \la{induced}
\frac{dN_g^{a}}{dzdk_\perp^2d\tau}=\frac{6\alpha_sP_a(z)k_\perp^4}{\pi (k_\perp^2+z^2m^2)^4} \frac{p\cdot u}{p_0}\hat{q}_{a} (x)\sin^2\frac{\tau-\tau_i}{2\tau_f},
\ea
where $m$ is the mass of the propagating parton, $z$  and $k_\perp$ are the energy fraction and transverse momentum of the radiated gluon, $P_a(z)$ the splitting function,  $\tau_f=2p_0z(1-z)/(k_\perp^2+z^2m^2)$ the gluon formation time, $\hat{q}_{a}(x)=\sum_{bcd}\rho_{b}(x)\int d\hat t q_\perp^2 d\sigma_{ab\rightarrow cd}/d\hat t $ the transverse momentum transfer squared per mean-free-path or  jet transport parameter in the local comoving frame, $\rho_{b}(x)$ is the parton density (including the degeneracy) and $\tau_i$ is the time of the last gluon emission. $\mu_D$ is used as an infrared cut-off for the gluon's energy. 

Within LBT, the probability of elastic scattering in each time step $\Delta \tau $ during the propagation of a parton is $P^a_{\rm el}=1-\text{exp}[- \Delta\tau \Gamma_a^{\rm el}(x)]$, where $\Gamma_a^{\rm el}\equiv \sum_{bcd} (p\cdot u/p_0)\rho_b(x)\sigma_{ab\rightarrow cd}$ is the elastic scattering rate.  The probability for inelastic process is $P^a_\mathrm{inel}=1-\exp[-\Delta\tau \Gamma_a^{\rm inel}(x)]$ where  $\Gamma_a^{\rm inel}=\int dz dk_\perp^2 (dN^a_g/dzdk_\perp^2d\tau)/(1+\delta_g^a)$ is the gluon radiation rate.
The total scattering probability $P^a_\mathrm{tot}=P^a_\mathrm{el}(1-P^a_\mathrm{inel}) +P^a_\mathrm{inel}$ can be split  into the probability for pure elastic scattering and the probability for inelastic scattering with at least one gluon radiation. Multiple gluon radiation is simulated by a Poisson distribution with the mean $\langle N^a_g \rangle=\Delta\tau\Gamma_a^{\rm inel}$. 

Since LBT is designed to study both jet propagation and j.i.m.e., all final partons after each scattering (jet shower partons, recoil medium partons and radiated gluons) will go through further scattering in the medium. To account for the back reaction in the Boltzmann transport, initial thermal parton $b$ in each scattering, denoted as ``negative'' partons with negative energy-momentum, are also transported according to the Boltzmann equation. They are part of the j.i.m.e. \cite{Wang:2013cia,Li:2010ts,He:2015pra} and their energies and momenta will be subtracted from all final observables. LBT has been employed to study successfully $\gamma$-jet modification, light and heavy flavor hadron suppression in heavy-ion collisions \cite{Wang:2013cia,Cao:2016gvr,Cao:2017hhk}.

{\it CoLBT-hydro model.}  
In LBT,  a hydrodynamic model provides information on the local temperature and fluid velocity of the bulk medium which evolves independently of the jet propagation.
Parton-medium interaction at all energy scales is described by pQCD and linear approximation ($\delta f\ll f$) is assumed which will break down when the j.i.m.e. becomes appreciable.  
To extend LBT beyond this region of applicability, we have developed CoLBT-hydro in which jet transport is coupled to hydrodynamic evolution of the bulk medium in real time and j.i.m.e. is also described by hydrodynamics. Such coupling is achieved through a source term in the hydrodynamic equation,  $\partial_\mu T^{\mu\nu}=j^\nu$,
\begin{eqnarray}
 j^\nu &=& \sum_i \frac{\theta(p^0_{\rm cut}-p_i\cdot u) dp_i^\nu/d\tau}{\tau(2\pi)^{3/2}{\sigma_r^2\sigma_{\eta_s}}} \nonumber \\
&\times& \exp\left[-\frac{(\vec x_\perp-\vec x_{\perp i})^2}{2\sigma_r^2}-\frac{(\eta_s-\eta_{si})^2}{2\sigma_{\eta_s}^2}\right],
\label{eq:source}
\end{eqnarray}
which is the energy-momentum deposition by soft ($p\cdot u<p^0_{\rm cut}$) and ``negative" partons ($p\cdot u<0$) from LBT with a Gaussian smearing. We employ the CCNU-LBNL viscous (CLVisc) code \cite{Pang:2014ipa,Pang:2012he} to solve the (3+1)D hydrodynamics with the above source term and a parametrized equation of state (EoS) s95p-v1\cite{Huovinen:2009yb}. CLVisc parallelizes Kurganov-Tadmor algorithm \cite{KURGANOV2000241} for space-time evolution of the bulk medium and Cooper-Frye particlization on GPU, using Open Computing Language (OpenCL). With massive amount of computing elements on GPUs and the Single Instruction Multiple Data (SIMD) vector operations on modern CPUs, CLVisc brings the best performance boosts so far to (3+1)D hydrodynamics on heterogeneous computing devices and makes event-by-event CoLBT-hydro simulations possible.

In CoLBT-hydro, both LBT and CLVisc  are formulated in the Milne coordinates $(\tau, \vec x_\perp, \eta_s)$ and are simulated in sync with each other. For each time step $\Delta\tau$, transport of jet shower partons are carried out according to LBT with local temperature and fluid velocity from CLVisc at $\tau$. Soft and ``negative" partons are removed from the list of partons  in LBT after each scattering and their energy-momentum contributes to the source term according to Eq.~(\ref{eq:source}) in the hydrodynamic evolution of the bulk medium. The updated local medium properties will be used in the transport of energetic partons $(p\cdot u > p^0_{\rm cut})$ within LBT for the next time step $\tau+\Delta\tau$. The initial energy-momentum density distributions for event-by-event CoLBT-hydro simulations are obtained from particles in A Multi-Phase Transport (AMPT) model \cite{Lin:2004en} with the same Gaussian smearing as in Eq.~(\ref{eq:source}) ($\sigma_r=0.6$ fm and $\sigma_{\eta_s}=0.6$). The normalization of the initial energy-momentum density, the initial time $\tau_0=0.4$ fm/$c$ and freeze-out temperature $T_{\rm f}=137$ MeV are fitted to reproduce experimental data on the final charged hadron rapidity and transverse momentum distributions \cite{Pang:2012he,Pang:2015zrq,Pang:2014pxa}. We employ the parton recombination model \cite{Han:2016uhh} developed within the JET Collaboration for hadronization of hard partons from LBT. The final hadron spectra from CoLBT-hydro include contributions from both LBT via parton recombination and CLVisc via Cooper-Frye freeze-out. The ideal version of CLVisc is used for most of our calculations. Detailed descriptions of the CoLBT-hydro model and the discussion of effect of viscosity will be given in a forthcoming publication.

\begin{figure}
\centerline{\includegraphics[width=8.5cm]{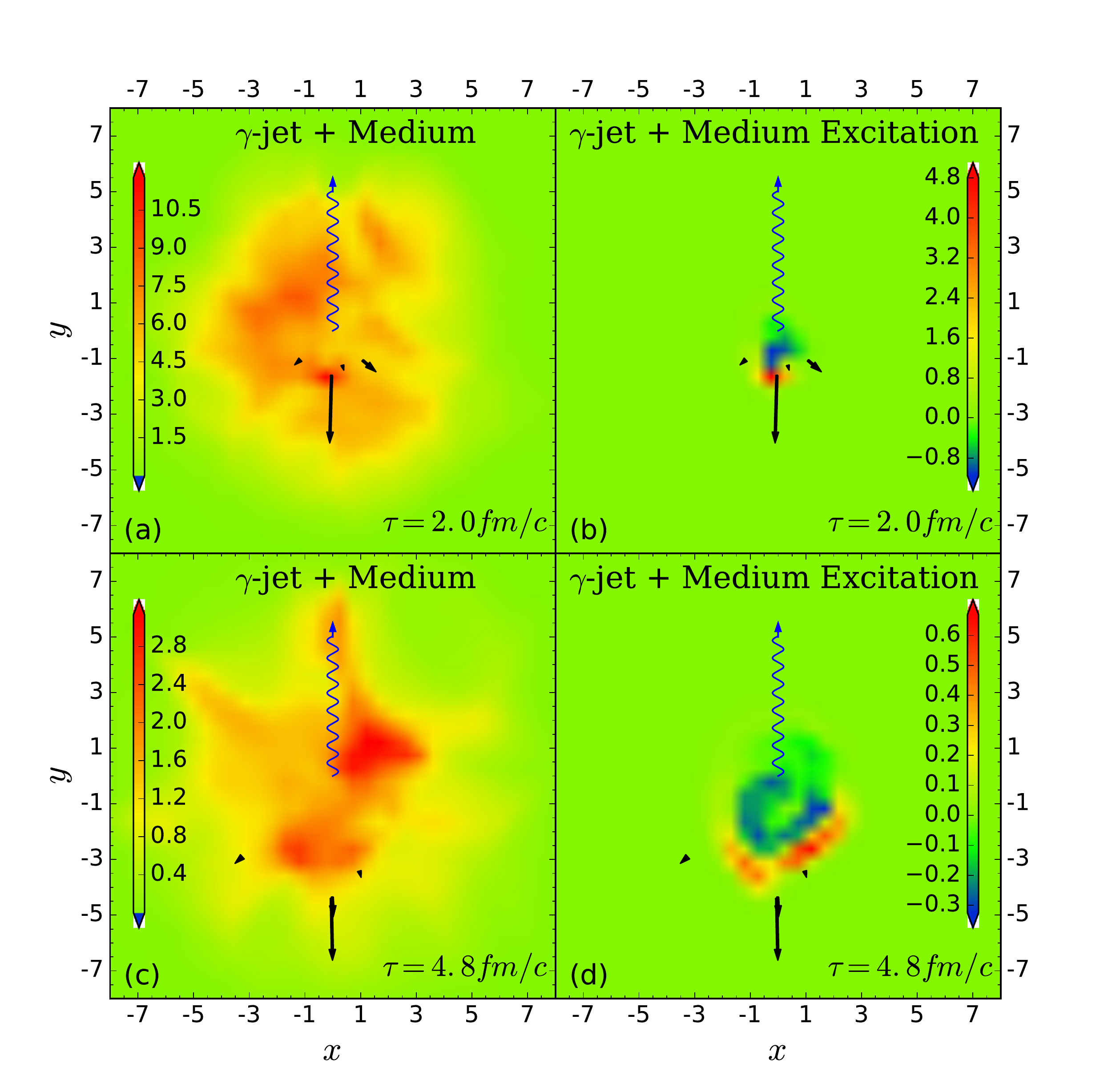}}
\vspace{-0.15in}
 \caption{Energy density (GeV/fm$^3$) and $\gamma$-jet evolution in the transverse plane at $\eta_s=0$, $\tau=2.0$ (a,b) and 4.8 fm/$c$ (c,d) in a 0-12\% central Au+Au collision at $\sqrt{s}=200$ AGeV. Straight (wavy) lines represent partons' (photon) momenta. Hydrodynamic background from the same event without $\gamma$-jet is subtracted in the right panels.}
 \label{movie}
\end{figure}

To illustrate jet transport and j.i.m.e. in CoLBT-hydro simulations we show in Fig.~\ref{movie} transverse distributions
of the energy density at two different time $\tau=2.0$ (upper panels) and 4.8 fm/$c$ (lower panels) in a 0-12\% central Au+Au collision at $\sqrt{s}=200$ AGeV with a $\gamma$-jet that is produced at the center of the overlap region.  The  (wavy) straight lines represent the momenta of ($\gamma$) hard jet shower partons. The left panel is from CoLBT-hydro with a $\gamma$-jet. The Mach-cone-like j.i.m.e. including the diffusion wake (depletion of energy density behind the jet) is clearly seen in the right panels where the same bulk medium evolution without the $\gamma$-jet is subtracted.

{\it $\gamma$-hadron correlation.} 
Modification of $\gamma$-hadron correlations has been proposed as a good probe of parton energy loss in QGP medium \cite{Wang:1996yh} since direct photons can be used to better measure the initial jet energy. We  carry out the first study of  jet quenching with CoLBT-hydro as well as j.i.m.e. through $\gamma$-hadron correlations in high-energy heavy-ion collisions.

We use Pythia8 \cite{Sjostrand:2007gs} to generate initial jet shower partons for $\gamma$-jet events in p+p collisions.
%with nuclear modified parton distributions that take into account of the isospin dependence in A+A collisions. 
These partons start to interact with the medium in CoLBT-hydro after their formation time $\tau_f=2p_0/p_T^2$ or the QGP formation time $\tau_0$ whichever later. The initial position of the $\gamma$-jet is sampled according to the spatial distribution of binary hard processes from the same AMPT event that provides the initial condition for the bulk medium evolution. The final hadron spectrum per $\gamma$ trigger, defined as the $\gamma$-triggered fragmentation function, \begin{equation}
D(z)=
\left. \frac{dN_h}{dz}\right|_{\rm LBT}+\left.\frac{dN_h}{dz}\right|_{\rm hydro}^{\rm w/jet}-\left.\frac{dN_h}{dz}\right|_{\rm hydro}^{\rm no/jet},
\end{equation}
$z=p_T^h/p_T^\gamma$, is the sum of hadron spectra from LBT and CLVsic in CoLBT-hydro minus the background from CLVisc with the same initial condition but without $\gamma$-jet.

\begin{figure}
\centerline{\includegraphics[width=8.5cm]{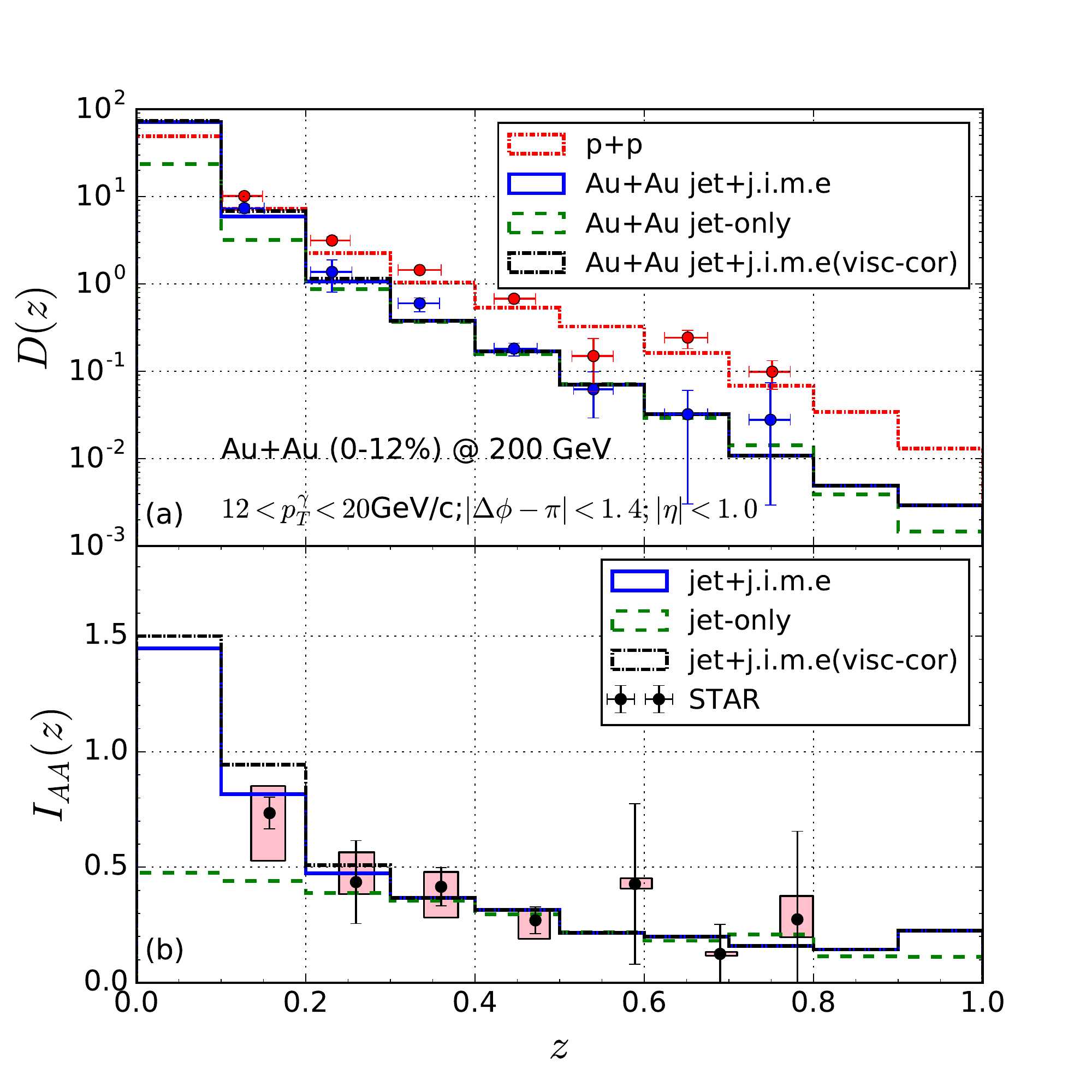}}
\vspace{-0.15in}
 \caption{ (a) $\gamma$-triggered jet fragmentation functions in p+p and 0-12\% Au+Au collisions at $\sqrt{s}=200$ AGeV and (b) the modification factor as compared to STAR data \cite{STAR:2016jdz}. Results without j.i.m.e. and with viscous correction (for $\eta/s$=0.16) are shown in dashed and dot-dashed lines, respectively.}
 \label{fig-star}
\end{figure}

Shown in Fig.~\ref{fig-star}(a) are CoLBT-hydro results for the $\gamma$-triggered fragmentation functions in p+p and 0-12\% central Au+Au collisions at $\sqrt{s}=200$ AGeV and (b) the corresponding modification factors $I_{AA}(z)=D_{AA}(z)/D_{pp}(z)$ for $12<p_T^\gamma<20$ GeV/$c$ within pseudo-rapidity $|\eta|<1$ and azimuthal angle $|\Delta \phi_{\gamma h}-\pi|<1.4$. A constant background in the hadron yield from CoLBT-hydro in p+p and Au+Au collisions is subtracted separately using the zero-yield-at-minimum (ZYAM) method similarly as in the experimental analyses. CoLBT-hydro describes well the STAR experimental data \cite{STAR:2016jdz} on suppression of leading hadrons at intermediate and large $z$ due to energy loss of hard partons within LBT. Soft hadrons at small $z<0.1$ are significantly enhanced due to contributions from j.i.m.e. as compared to that without (also excluding recoil thermal partons in LBT). The only parameter that controls parton energy loss in LBT is the strong coupling constant which we choose $\alpha_{\rm s}=0.3$ to fit the STAR data. The cut-off parameter is set at $p^0_{\rm cut}=2$  GeV for soft partons that contribute to the source term for CLVisc. The final combined spectra from LBT and CLVisc are not sensitive to $p^0_{\rm cut}$ within the range $1<p^0_{\rm cut}<4$ GeV. Though the effect of viscosity can be significant for  hadron spectra from the bulk \cite{Ryu:2015vwa} and j.i.m.e. at large $p_T$, viscous correction to the final fragmentation function is only sizable (about 10-20\% for $\eta/s=0.16$) at low $z$ where j.i.m.e. contribution dominates as shown by the dot-dashed lines.

\begin{figure}
\vspace{-0.20in}
\centerline{\includegraphics[width=8.5cm]{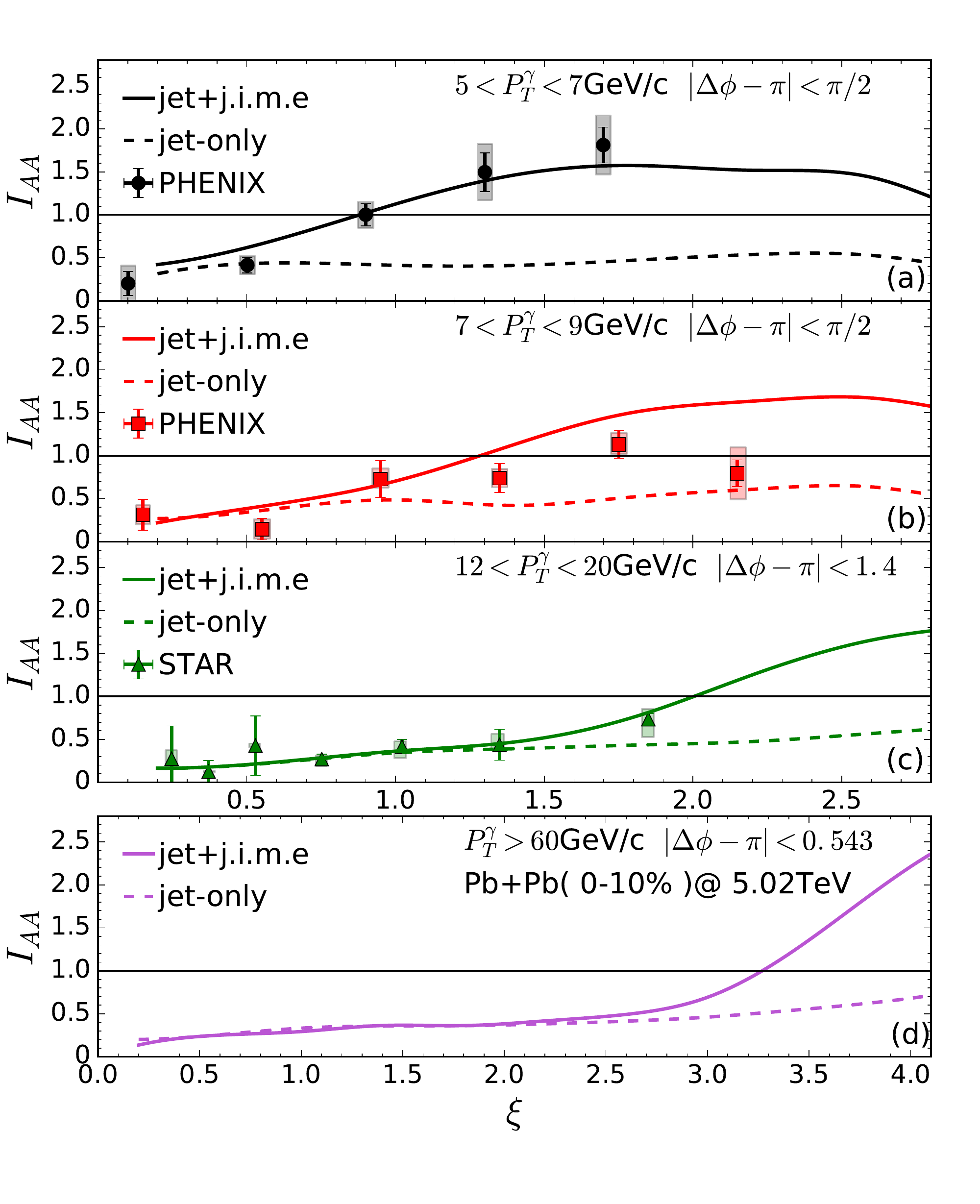}}
\vspace{-0.30in}
 \caption{Modification factor for $\gamma$-hadron correlation as a function of $\xi=\log(1/z)$ for $|\eta_{h,\gamma}|<0.35$ and different $p_T^\gamma$ in 0-40\% and 0-12\% Au+Au collisions at $\sqrt{s}=200 $ AGeV with (solid) and without j.i.m.e (dashed) as compared to the STAR \cite{STAR:2016jdz} and preliminary PHENIX data \cite{Ge:2017irb}.}
 \label{fig-phenix}
\end{figure}

To exam j.i.m.e. in detail, we show in Fig.~\ref{fig-phenix}(a)-(c) the modification factor $I_{AA}$ for $\gamma$-hadron correlation with (solid) and without j.i.m.e. (dashed lines) as a function of $\xi=\log(1/z)$ in 0-40\% and 0-12\% Au+Au collisions at $\sqrt{s}=200$ AGeV for different ranges of $p_T^\gamma$ within $|\eta|<0.35$ and $|\Delta \phi_{\gamma h}-\pi|<\pi/2$  as compared to experimental data \cite{STAR:2016jdz,Ge:2017irb}.  Jet-medium interaction in CoLBT-hydro leads to the suppression of hadrons at small $\xi$ as well as enhancement at large $\xi$ that are consistent with STAR \cite{STAR:2016jdz} and preliminary PHENIX data  \cite{Ge:2017irb}. Since soft hadrons from  j.i.m.e. carry an average thermal energy that is independent of the jet energy, a unique feature of the CoLBT-hydro results is that the onset of soft hadron enhancement ($I_{AA}\ge 1$) due to j.i.m.e. occurs at a constant $p^h_T\sim 2 $ GeV/$c$ or $\xi=\ln(p_T^\gamma/p_T^h)$ that increases logarithmically with $p_T^\gamma$. We also show in Fig.~\ref{fig-phenix}(d) our prediction on $I_{AA}$ for $p_T\gamma>60$ GeV in central Pb+Pb collisions at the LHC energy $\sqrt{s}=5.02$ TeV.

\begin{figure}
\centerline{\includegraphics[width=8.5cm]{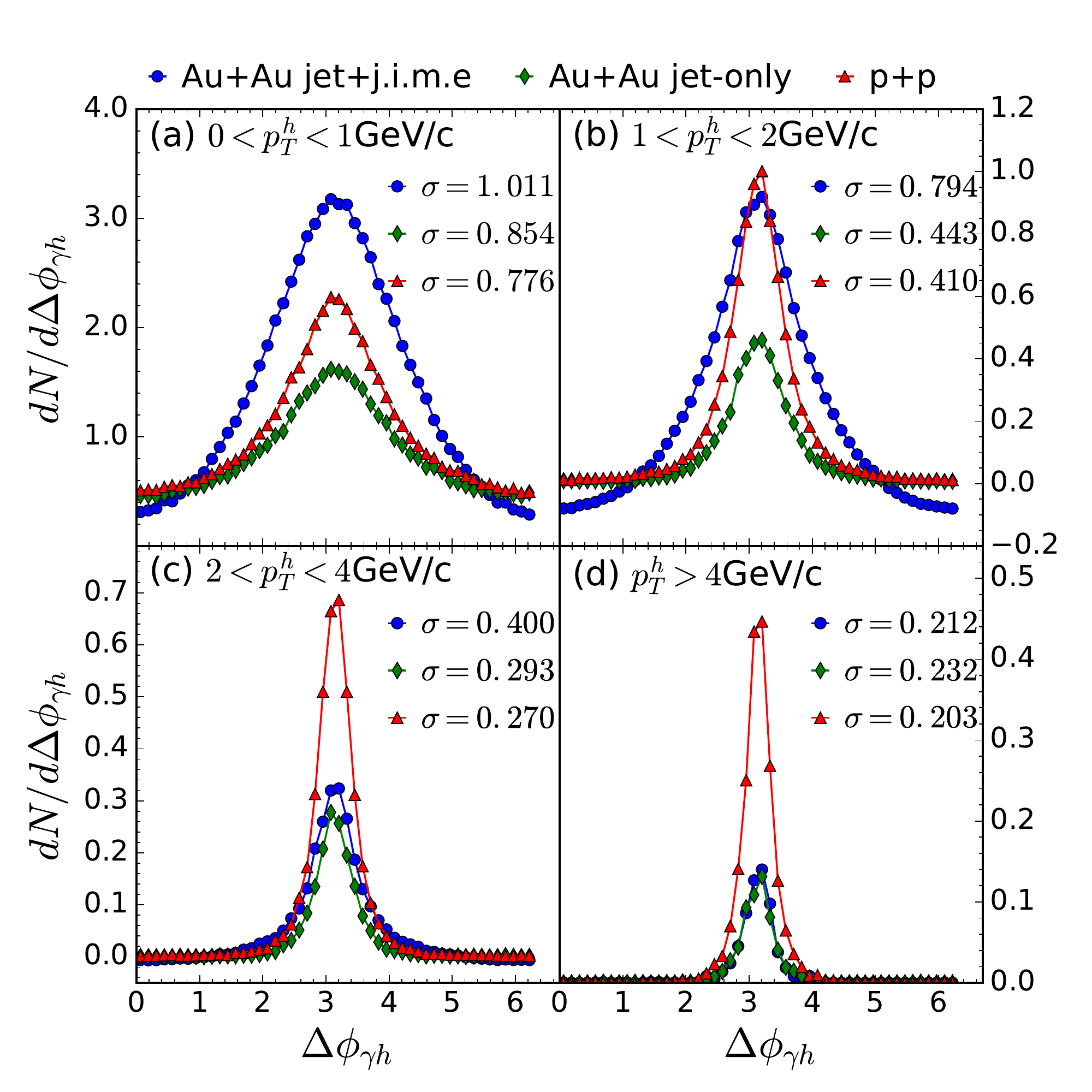}}
\vspace{-0.10in}
 \caption{$\gamma$-hadron azimuthal correlation for $|\eta_{h,\gamma}|<1.0$,  $12<p_T^\gamma<20$ GeV/$c$ and different $p_T^h$ in p+p (triangle) and 0-12\% central Au+Au collisions with (circle) and without (diamond) j.i.m.e. at $\sqrt{s}=200$ AGeV. The half width $\sigma$ is obtained via a Gaussian fit within $|\Delta\phi_{h\gamma}-\pi|<1.4$.}
 \label{dndphi}
\end{figure}

To illustrate the angular structure of j.i.m.e., we plot in Fig.~\ref{dndphi} $\gamma$-hadron correlation as a function of $\Delta\phi_{\gamma h}$ in p+p and 0-12\% central Au+Au collisions at $\sqrt{s}=200$ AGeV (without ZYAM background subtraction).  Large $p_T$ hadron yields from $\gamma$-jet in Au+Au are suppressed but the width of their angular distributions remain approximately unchanged from p+p. The angular distributions for the enhanced soft hadrons in Au+Au are, however, significantly broadened. The enhancement due to j.i.m.e. occurs both in the small $|\Delta \phi_{\gamma h}-\pi|<\pi/6$ and large azimuthal angle $\pi/3<|\Delta \phi_{\gamma h}-\pi|<\pi/2$ region relative to the jet direction. The most interesting feature in the angular distribution of soft hadrons is the depletion of soft hadrons in the  $\gamma$ direction due to the diffusion wake left behind by the jet in QGP.

{\it Summary.} 
We have developed the state of art CoLBT-hydro for co-current and event-by-event simulations of jet propagation and hydrodynamic evolution of the bulk medium including j.i.m.e.. We carried out the first study with CoLBT-hydro of the medium modification of $\gamma$-hadron correlations in heavy-ion collisions at RHIC. CoLBT-hydro describes well the suppression of leading hadrons due to parton energy loss and predicts an enhancement of soft hadrons due to j.i.m.e.. The onset of soft hadron enhancement at a constant $p_T^h$ with broadened angular distribution
%relative to the jet direction 
and depletion of soft hadrons in the $\gamma$ direction are two unique features of j.i.m.e. that are different from the parton cascade picture \cite{Blaizot:2013hx,Fister:2014zxa}.
%The onset of the enhancement is found to occur at fixed $p_T^h$ that is independent of $\gamma$ or jet energy, indicating the thermal nature of hadrons from j.i.m.e. which also leads to broadened azimuthal distributions along the jet direction and depletion of soft hadrons in the $\gamma$ direction due to the diffusion wake. rap
Experimental studies of these effects at RHIC ( sPHENIX) and LHC should provide a new window into jet tomography of QGP in high-energy heavy-ion collisions.
% More detailed study of j.i.m.e. can also provide information about the transport properties of QGP in high-energy heavy-ion collisions. 

\begin{acknowledgments}
 This work is supported by the NSFC under grant No. 11521064, MOST of China under Projects No. 2014CB845404,
 %MOST of China under grant No. 2014DFG02050, 
 NSF 
 %within the JETSCAPE Collaboration and 
 under grant No. ACI-1550228,
 % and ACI-1550300, 
 U.S. DOE under Contract Nos. DE-AC02-05CH11231 and DE-SC0013460. Computations are performed at the Green Cube at GSI
and GPU workstations at CCNU.
\end{acknowledgments}


\end{document}